\documentclass[aps,prb,twocolumn,superscriptaddress,groupedaddress]{revtex4}
\usepackage{hyperref}
\usepackage{bm}
\usepackage{graphicx}
\usepackage{amsmath}
\usepackage{sidecap}
\usepackage{amsfonts}

\usepackage{color}

\begin{document}
	\title{Quenched randomness, thermal fluctuations and reentrant superconductivity: application to UTe$_2$}
	\author{Yue Yu}
	\affiliation{Department of Physics, Stanford University, 476 Lomita Mall, Stanford, CA 94305}
	\author{S. Raghu}
	\affiliation{Department of Physics, Stanford University, 476 Lomita Mall, Stanford, CA 94305}
	\affiliation{Stanford Institute for Materials and Energy Sciences, SLAC National Accelerator Laboratory,
		Menlo Park, CA 94025, USA}
	\begin{abstract}
		Reentrant superconductivity has been observed in the candidate spin-triplet superconductor UTe$_2$ as a function of the magnetic field applied along the hard axis. Resistivity measurements have shown, a broadened superconducting transition appears near the minimal $T_c$, highlighting the importance of superconducting fluctuations in this regime. We present a phenomenological study assuming a field-driven first-order transition between two superconducting states. We show that with quenched randomness,  inhomogeneity-enhanced superconducting fluctuations near the transition could naturally account for both the reentrant superconductivity as well as the broadened superconducting transition.
	\end{abstract}
	\maketitle
	
	\section{introduction}
	
	Magnetic fields are often detrimental to superconductivity.  Spin singlet superconductors are destroyed by magnetic fields both by the Zeeman effect and through the orbital motion of charge carriers.  Spin triplet superconductors can be more robust since the Zeeman effect does not break apart triplet pairs for certain field orientations.  Nevertheless, they are not immune to the orbital depairing effects in a generic three-dimensional system.  In striking contrast to these expectations, field-induced reentrant behavior, where superconductivity strengthens and even emerges with a magnetic field, { has been observed in several heavy fermion systems CeRh$_2$As$_2$ \cite {Khim2021}, UGe$_2$ \cite{sheikin2001}, URhGe \cite{levy2005}, UCoGe \cite{aoki2009} and UTe$_2$ \cite{knebel2019},  and the magic-angle twisted trilayer graphene \cite{cao2021}.  }
	
	In the case of UGe$_2$, URhGe, and UCoGe superconductivity directly coexists with ferromagnetism and reentrant behavior is associated with a spin-triplet condensate with magnetically ordered cooper pairs.  In these systems, the conjectured source of reentrance is that the field tunes the system towards a quantum phase transition \cite{Belitz1999,miyake2009,nakamura2017}, whose associated fluctuations enhance the pairing interaction in the spin-triplet channel and thus result in an increased transition temperature.  \footnote{Alternatively the magnetic field tunes the system across a Lifshitz transition where the Fermi surface topology changes, resulting in sharp changes to the superconducting transition temperature with field}  By contrast, superconductivity does not coexist with ferromagnetism in UTe$_2$ \cite{ran2019} and the idea that a proximate quantum phase transition is responsible for reentrance appears to be less tenable since the observed quantum phase transition in a field appears to be first-order \cite{miyake2019,knebel2019}.  If the enhanced magnetic fluctuation is due to a weakly first-order metamagnetic transition, a superconducting dome is expected, as in URhGe\cite{levy2005} and UCoGe\cite{aoki2009}. The lack of a superconducting dome in UTe$_2$ and a drastic change in normal state resistivity question the validity of this possibility.

	We provide here an alternate mechanism for reentrance in UTe$_2$ which does not involve quantum critical ferromagnetic fluctuations.  Indeed, predominantly incommensurate antiferromagnetic fluctuations have been observed in this system\cite{Duan2020, Knafo2021}.  UTe$_2$ also shows multiple superconducting phases especially as a function of pressure\cite{Ran2020,thomas2021}.  Further evidence for multiple phases comes from a recent Kerr effect study suggesting the interesting possibility of spin-triplet superconductivity with broken time-reversal symmetry.  These studies suggest that the competition between multiple superconducting phases is an important underlying feature.  We show here how these multiple phases can give rise to reentrant superconductivity. 
	
	We postulate that in the absence of disorder, the magnetic field drives the system across a first-order transition separating two distinct superconducting ground states (Fig 1).  In a mean-field approximation, the first-order boundary can terminate at a bicritical point.  With quenched randomness, thermal fluctuations are enhanced near the bicritical point, which suppresses the transition temperature in its vicinity.  The reentrance naturally occurs in this scenario as a consequence of the non-monotonic magnitude of such thermal fluctuations as a function of the magnetic field.  
	
	Our scenario is inspired by recent experimental studies, which have revealed the sharpness of the superconducting transition as a function of the field.  These studies conclude that the resistive transition remains sharp for small and large fields, but for intermediate fields, the resistive transition is broadened, indicating enhanced thermal fluctuations.  Our scenario naturally leads to reentrance and has several sharp testable experimental consequences.

	\section{Proposed Phase diagram}\label{S1}
\begin{figure}[h]
		\centering
		\includegraphics[width=7cm]{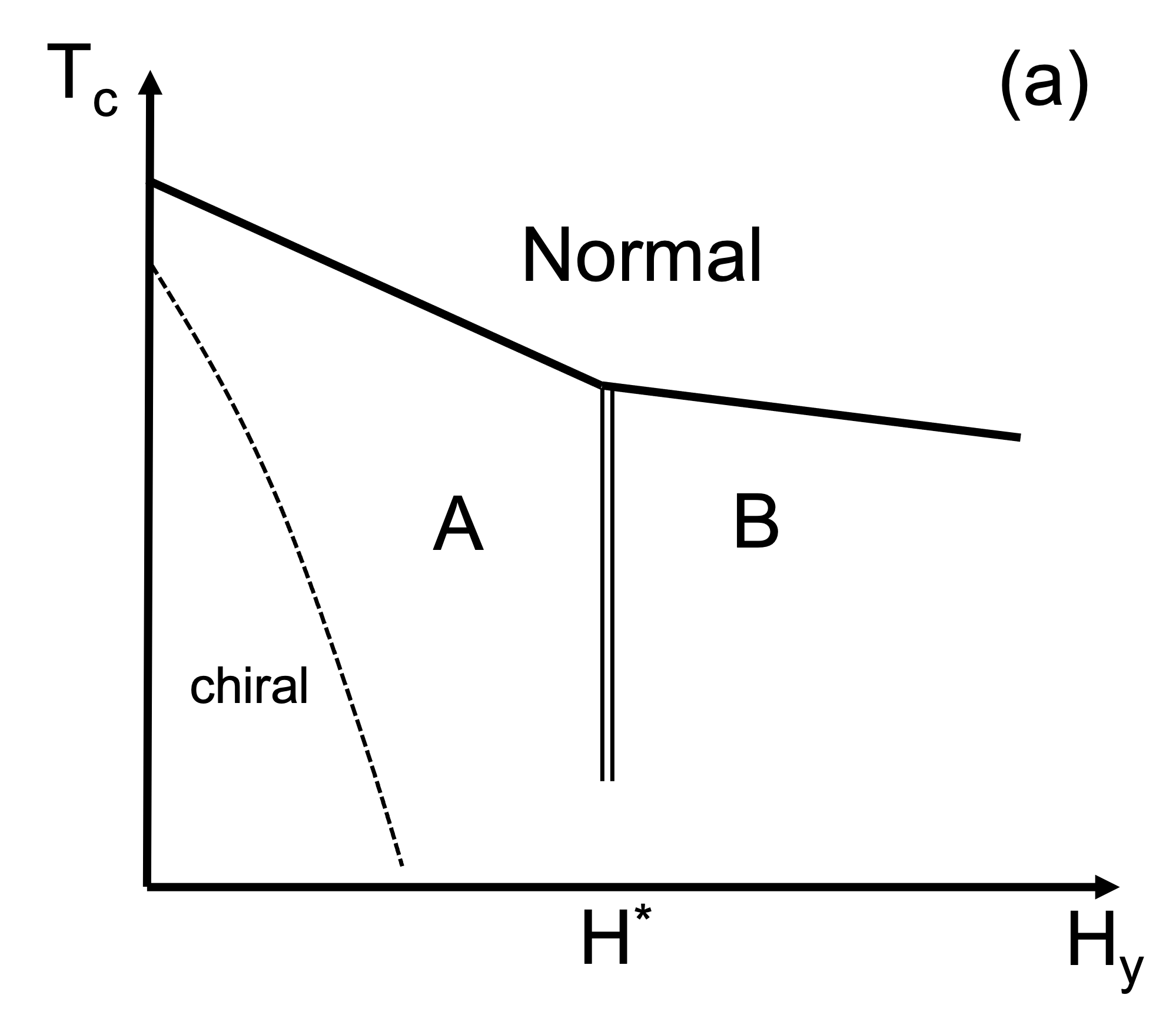}
		\includegraphics[width=7cm]{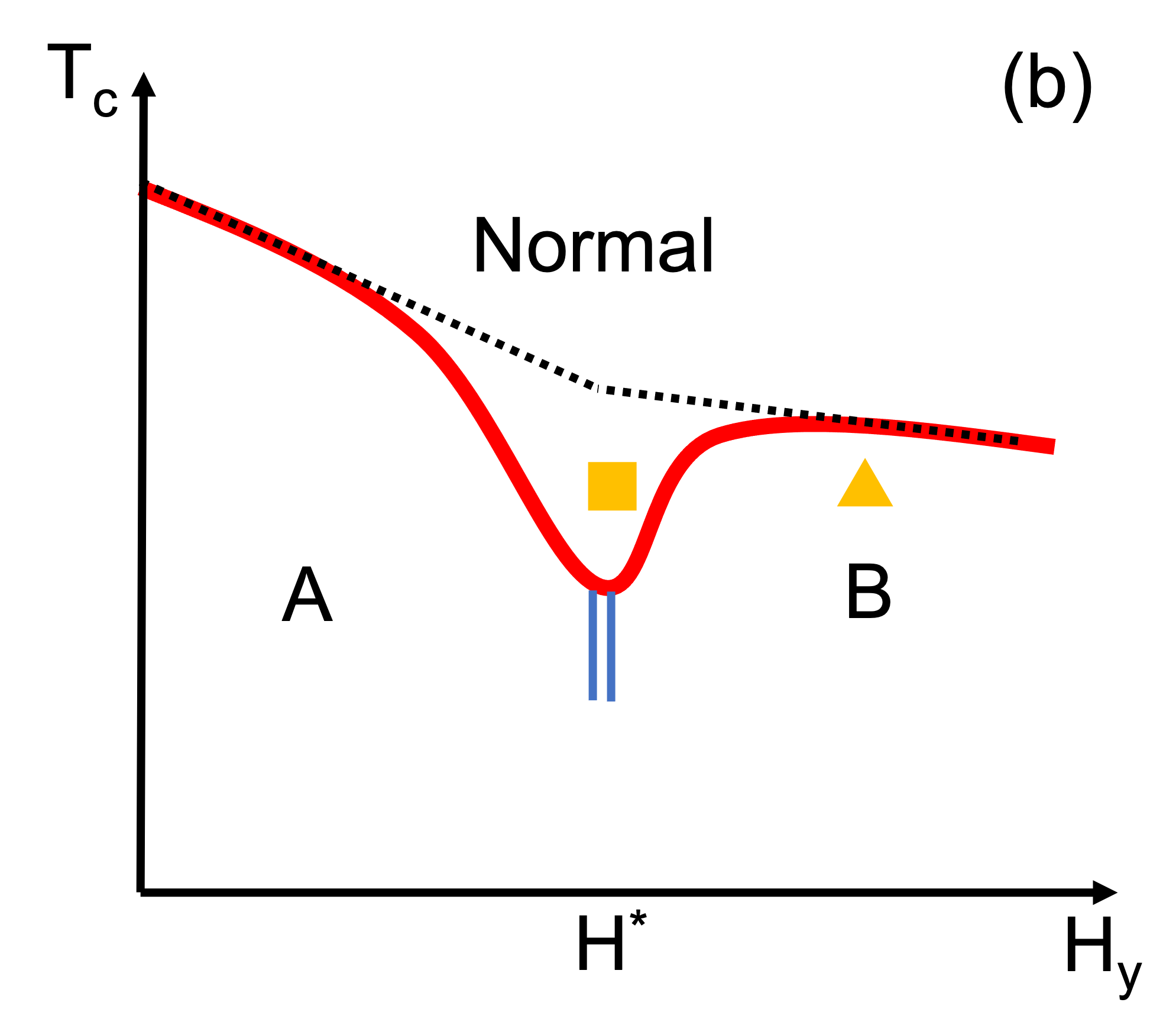}
		\caption{(Top)Mean-field phase diagram in the clean system. (Bottom)Phase diagrams under fluctuation with random-$T_c$ disorder. Double solid lines are first-order transitions, and single solid lines are continuous transitions. Dotted lines are cross-over. Descriptions of the transition lines can be found in the main text. Classical Monte Carlo simulations are applied on the orange points.}
		\label{f1} 
	\end{figure}

	In this section, we propose a mean-field phase diagram for the clean system (Fig.\ref{f1}(a)) and obtain fluctuational corrections in the presence of quenched randomness. The crucial assumption is the first-order transition between two superconducting phases A and B at $H=H^*$, denoted in double solid lines.  These phases might, for instance, be related to the multiple superconducting phases observed with applied pressure, but for the present discussion, their specific properties are unimportant.  A magnetic field applied along a crystalline $y$-axis suppresses both phases, by orbital depairing, and the superconducting $T_c$ monotonically decreases as a function of field. To drive the first-order transition, the magnetic field needs to suppress phase A more strongly than phase B, which amounts to stating that the superconducting coherence length of phase B is short compared to that of phase A.  Further details on the magnetic field suppression of the two phases are provided in Sec. \ref{S2}. 
	%More detailed discussions on the suppression and symmetries of the two phases are in Sec.\ref{S2}.  For simplicity, the possible chiral phase near zero-field is assumed to be fully suppressed before $H=H^*$, therefore is not relevant for the following analysis. A scenario with weaker suppression of the chiral phase can be found in Sec.\ref{S3}.
	
	We next include weak quenched randomness and obtain fluctuational corrections to the superconducting $T_c$. A schematic phase diagram can be found in Fig.\ref{f1}(b). A generic type of disorder is a random-$T_c$, which randomly increases or decreases the local critical temperature of orders A and B, therefore affecting the local preference of orders. 
	
	Below the mean-field $T_c$, and far away from the first-order transition, the system is mainly occupied by the dominant order. Due to the disorder, the subdominant order can appear as rare superconducting regions. The superconducting $T_c$ here is determined primarily by the long-range ordering of the dominant order, therefore, is similar to the mean-field $T_c$ in the clean system. The resistive transitions in these regions would therefore remain sharp, and the effects of fluctuations weak.%The fluctuation effect is weak, and the superconducting transition is sharp.

	By contrast, closer to the first-order transition, patches of the subdominant order are more prevalent. Superconducting fluctuations are enhanced by the resulting spatial inhomogeneity, resulting in a substantial lowering of $T_c$ (thick red line) relative to the mean-field value. The mean-field critical temperature now characterizes the local ordering within superconducting regions of the dominant order. In this region, therefore, the resistivity deviates from its normal state value at this cross-over temperature $T_{onset}$ (black dotted line), which would appear as a broadened superconducting transition. 
	
	The broadest transition and $T_c$ minimum, therefore, appear at the same magnetic field $H=H^*$. As a function of the magnetic field, the increase in $T_c$ from its minimum to the maximum is smaller or equal to the broadened transition $\Delta{T}=T_{onset}-T_c$ at $H=H^*$. From the resistivity data, these two quantities are both around $0.5K$.  The possibility of additional transitions in this picture within the superconducting phases will be discussed in the next section.
	
	\section{Possible phases for A \& B}\label{S2}
	The discussion above made no explicit reference to a particular set of phases; our only assumption was the existence of a first-order transition and a bicritical point separating two distinct superconducting phases in the absence of disorder.  The effect of disorder then enhances thermal fluctuations near the bicritical point.  In this section, we discuss the possible choices for phases A and B.  
	%	We will firstly discuss the possible choices for phases A and B. 
	
\begin{figure}[h]
		\centering
		\includegraphics[height=5cm]{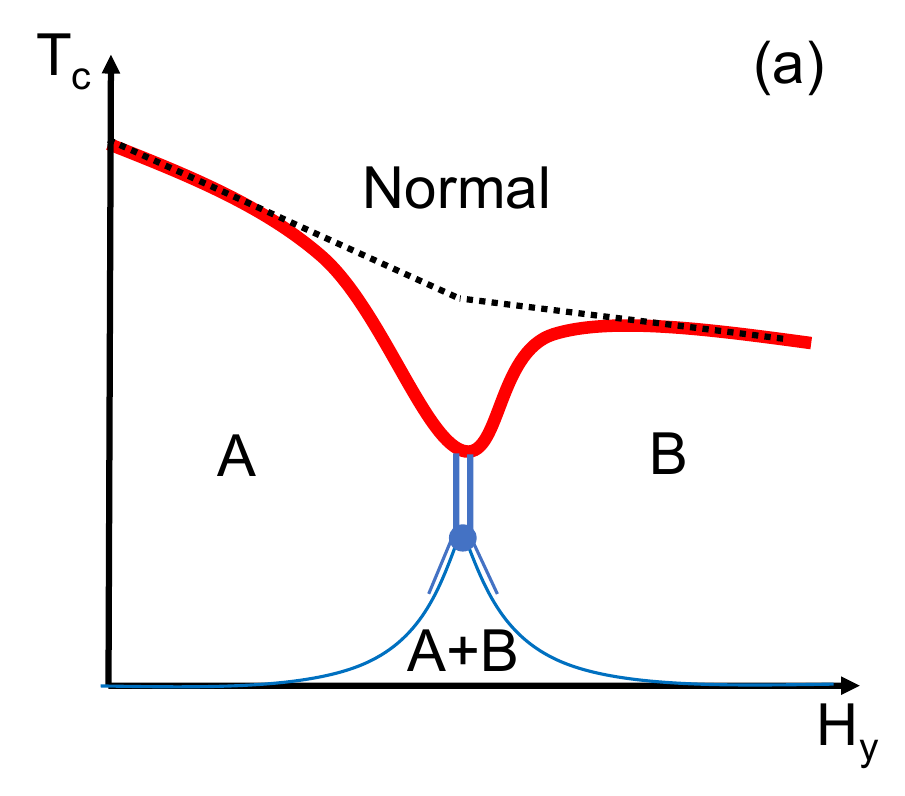}
		\includegraphics[height=5cm]{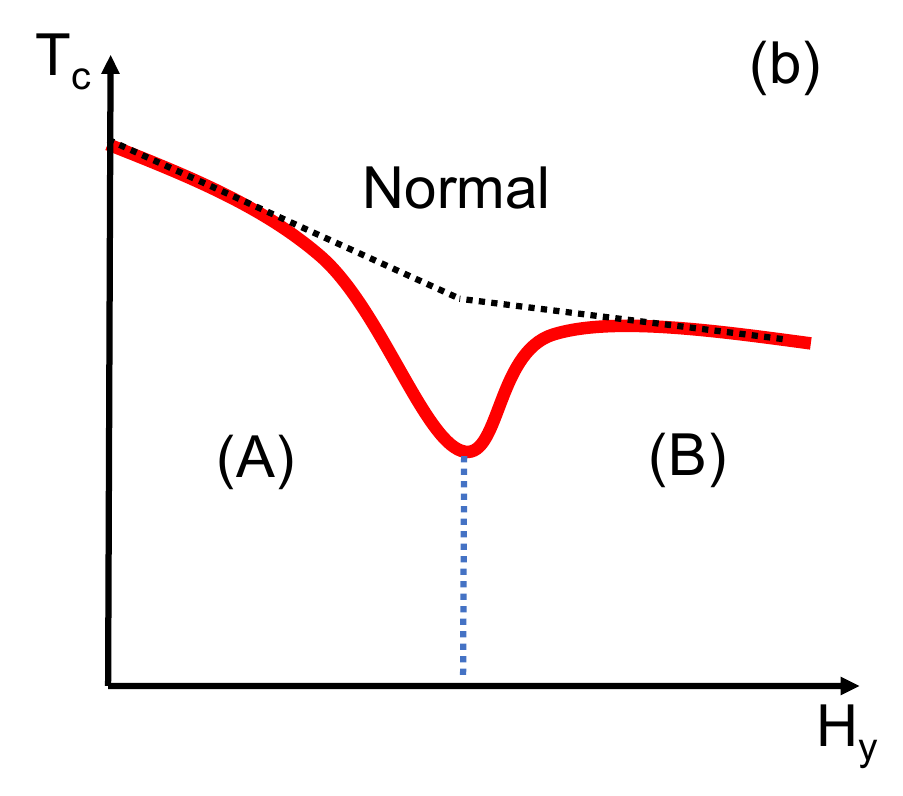}
		\includegraphics[height=5cm]{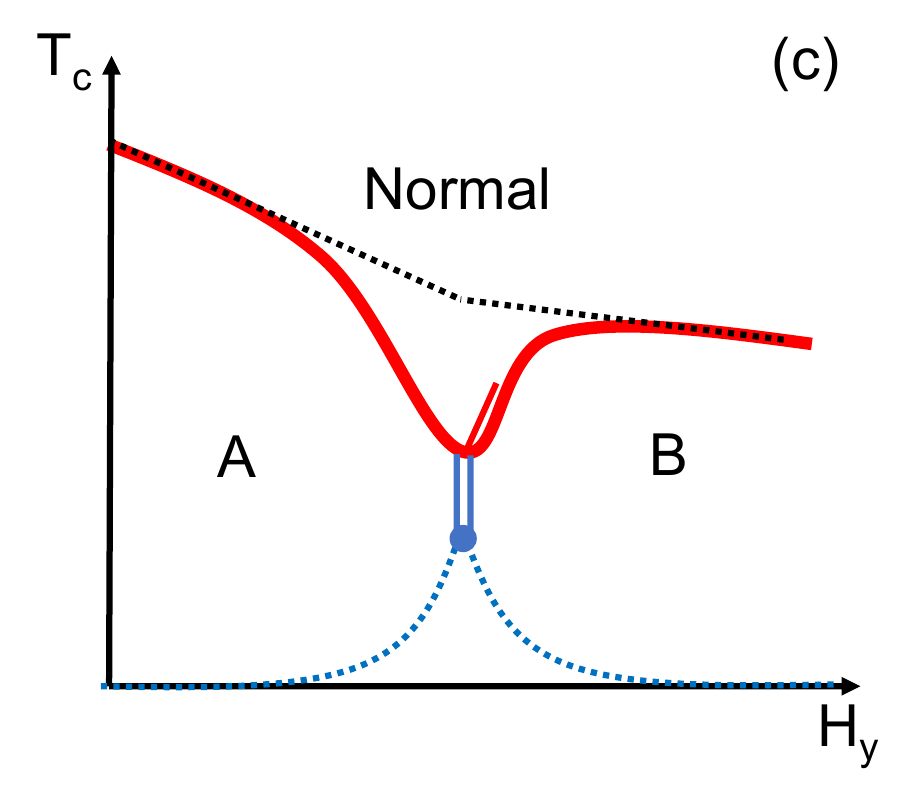}
		\caption{(Top)A and B are distinguished by pairing symmetries under $C_{2h}$. (Middle) A and B are distinguished by vortex lattice structures in the clean system. (Bottom)A and B have the same symmetries. Double solid lines are first-order transitions, and single solid lines are continuous transitions. Dotted lines are cross-over.  }
		\label{f2} 
	\end{figure}

	%They
	The simplest scenario is one in which the two phases
	have different pairing symmetries (Fig.\ref{f2}(a)). Such a scenario is corroborated by recent Kerr effect measurements (at zero $H_y$), which have revealed two superconducting transitions below which a non-zero magnetization along the $c$ axis develops. 
	In an orthorhombic system, this requires two nearly degenerate irreducible representations. These two states in turn still belong to different irreducible representations with non-zero $H_y$.  As an example, the two states could be $(\bm d_1(\bm k),\bm d_2(\bm k)) =(k_x \hat z,k_y \hat z)$ or $(k_x \hat x, k_y \hat x)$. The state $\bm d_1$ breaks reflection symmetry $y \rightarrow -y$ and preserves the $180^\circ$ rotation along y-axis; while the state $\bm d_2$ behaves oppositely. 
	It is natural to consider a field-driven transition between these two phases. Since UTe$_2$ is a 3D crystal without dramatic resistivity anisotropy, we would expect that the first-order transition is preserved under weak random-$T_c$ disorder. In the superconducting phase, subdominant local patches could develop long-range ordering, leading to a second transition to the coexistent phase. In the coexistent phase, the relative phase between the two orders could be either $0$ (``A+B") or $90^\circ$ (``A+iB"). 
	
	A second scenario is one in which phases A and B in the clean system correspond to distinct vortex lattice structures (Fig.\ref{f2}(b)). With quenched randomness, the vortex lattice order parameter maps on to the XY model in a random field, whose lower critical dimension is 4; %In this case, the order parameters of the vortex lattice suffer from the random-field disorder, which leads to 
	dislocations 
	%and 
	always destroy the vortex lattice phase in a 3-dimensional system. Thus, with the disorder, the phases themselves cease to have long-range order, and the first-order transition separating them is destroyed.  % isn't a sharp distinction between the two phases, and the first-order transition is destroyed. 
	With weak random fields, dislocations are rare and the majority of the system is uniform. 
	In the limit of weak random fields, the length scale beyond which long-range vortex crystalline order is destroyed is large, and short-range correlations of the crystal will persist.  In this case, 
	%All the previous descriptions of random-$T_c$ disorder will apply in those uniform regions. 
	many of the conclusions above, namely strong fluctuational corrections near the now avoided bicritical point survive.  Indeed, the random fields lead to even stronger superconducting inhomogeneity near what was the first-order transition, and once again, $T_c$ suppression in this region will result in reentrant superconductivity.  The enhanced inhomogeneity enables the subdominant order to survive more easily in the vortex cores of the dominant order.  
	%Typically, there will be enhanced inhomogeneity near the original first-order transition. Random-field disorder, in fact, also leads to enhanced inhomogeneity near the original first-order phase transition, where the subdominant order finds it easier to survive in the dislocation cores of the dominant order.
	
	In the third scenario, the two phases have the same symmetry (Fig.\ref{f2}(c)).  In this case, the first-order transition in the clean limit is precisely analogous to the liquid-gas transition. With quenched disorder, inhomogeneous superconductivity would result in a coexistent region, where both A and B develop long-range ordering. This implies that the first-order transition may end at a critical endpoint, followed by cross-over lines. { Note that the original bicritical point is in general not allowed in this scenario.}

	\section{magnetic suppression}	
	We next invoke the fact that the reentrant behavior in UTe$_2$ occurs as a function of the magnetic field.  Since the magnetic field suppresses pairing both via the Zeeman and orbital effects, we consider each case separately.  
	%Next, we will discuss the nature of the field-driven transition. Magnetic field suppresses superconductivity through Pauli paramagnetism effect and orbital effect. 
	Studies of magnetic field angle dependence\cite{knebel2019} have revealed that the orbital depairing is more relevant to UTe$_2$, at least for weak magnetic fields.  We, therefore, concentrate on the orbital effect.  For the sake of completeness, we also provide a discussion of the case where the Zeeman effect primarily destroys pairing.
	%	Under a magnetic field slightly away from the b-axis, the reentrant superconductivity can be fully suppressed\cite{knebel2019}, while the shape of $T_c$ at the small field is approximately unchanged. This part of the suppression can be fitted with the standard orbital-limited effect. Therefore, orbital suppression is crucial in this system. Pauli-limited suppression is known to have a slower suppression in $T_c$ and we are unsure whether or not it plays an important role in the phase transitions between phase A and B.
	
	If the transition is driven by the orbital effect, we would require different coherence lengths for phases A and B. For phase A, it needs to have a smaller critical field, thus a longer coherence length. For phase B, it needs to have a shorter coherence length. If the transition is driven by the Pauli effect, we would require a difference in the orientation of the d-vector. Phase A needs to have a d-vector, that is more parallel to the field. d-vector of phase B needs to be more perpendicular to the field. 
	
	{ For completeness, both Pauli and orbital-driven transitions will be described in the next section in the classical Monte Carlo simulation. For UTe$_2$, orbital suppression is more relevant, since the initial drop in $T_c$ under small magnetic field is linear in terms of the field strength, while that drop is quadratic within the Pauli-limited theory.}

	\section{Monte Carlo simulation}\label{S3}
	In this section, we will apply classical Monte Carlo simulation to investigate the thermal fluctuations enhanced by spatial inhomogeneity.  For simplicity, and purpose of illustration, we will first consider the case where the magnetic field couples solely to the spin of the condensate, resulting in Pauli suppression.  We then consider the complementary case where the field couples via the vector potential and induces orbital motion of charges.  
	
	We would like to compare the correlations of the superconducting order parameters in the neighborhood of the orange square and triangle points in the phase diagram in Fig.\ref{f1}b. The square point is taken from the region of enhanced thermal fluctuations near the original bicritical point. In this region, correlations are long-ranged in the clean limit but decay exponentially in the presence of disorder, as is clearly seen in Fig.\ref{f3}. By contrast, the region marked by the triangle in Fig.\ref{f1}b, which is at a higher magnetic field, but at the same temperature as the region marked by the square, has long-ranged correlations both in the clean and disordered cases. By comparing these two regions, we clearly see that inhomogeneity induces reentrant superconductivity.  
	
	In the following Monte Carlo simulations, fluctuation effect is purely from phase fluctuation in A and B patches, while $T_c$ is governed by Josephson couplings. Fluctuation of magnetic field, which may be crucial to suppress $T_c$, is not included in the calculation.

	\subsection{purely Pauli paramagnetism effect}
	We first consider the phase transition driven by purely Pauli paramagnetism effect. We consider a toy model under tetragonal symmetry, with $(A,B)$ identified as $(p_y,p_x)$ states. The analysis for other symmetries is similar. The free energy density is \cite{Sigrist1991}:
	
	\begin{equation}
		\begin{split}
			&f({\bf r})=-\frac{u_A({\bf r},H,T)}{2}|\Delta_A|^2-\frac{u_B({\bf r},H,T)}{2}|\Delta_B|^2\\
			&+b_1(|\Delta_A|^2+|\Delta_B|^2)^2+\frac{b_2}{2}\left[\Delta_A^2(\Delta_B^*)^2+c.c.\right]\\
			&+b_3|\Delta_A|^2|\Delta_B|^2+\frac{K_1}{2}(|\partial_x\Delta_B|^2+|\partial_y\Delta_A|^2)\\
			&+\frac{K_2}{2}(|\partial_x\Delta_A|^2+|\partial_y\Delta_B|^2)\\
			&+\frac{K_3}{2}\left[(\partial_x\Delta_B)^*(\partial_y\Delta_A)+c.c.\right]\\
			&+\frac{K_4}{2}\left[(\partial_x\Delta_A)^*(\partial_y\Delta_B)+c.c.\right]\\
			&+\frac{K_5}{2}(|\partial_z\Delta_A|^2+|\partial_z\Delta_B|^2)
		\end{split}
		\label{eqGL}
	\end{equation}
	In the Landau theory without the disorder, fluctuation, magnetic field, or competition, the quadratic coefficients can be related to the bare critical temperature of the two order parameters by $u=a(T_c-T)$. Under strong competition, the dominant order needs to have a larger quadratic coefficient.
	We now drop the temperature-dependence since we are focusing on the two points at the same temperature. Pauli paramagnetism effect suppresses orderings by reducing $u_A$ and $u_B$:  
	\begin{equation}
		\begin{split}
			&u_A({\bf r},H,T)=\overline{u_A(H)}+\delta{u_A}({\bf r})\\
			&u_B({\bf r},H,T)=\overline{u_B(H)}+\delta{u_B}({\bf r}).
		\end{split}
	\end{equation}
	Random $T_c$ disorder $\delta{u_{x,y}}({\bf r})$ is taken from independently uniformly distributed interval $[-\Delta{u},\Delta{u}]$. 
	
	\begin{figure}[h]
		\centering
		\includegraphics[width=6cm]{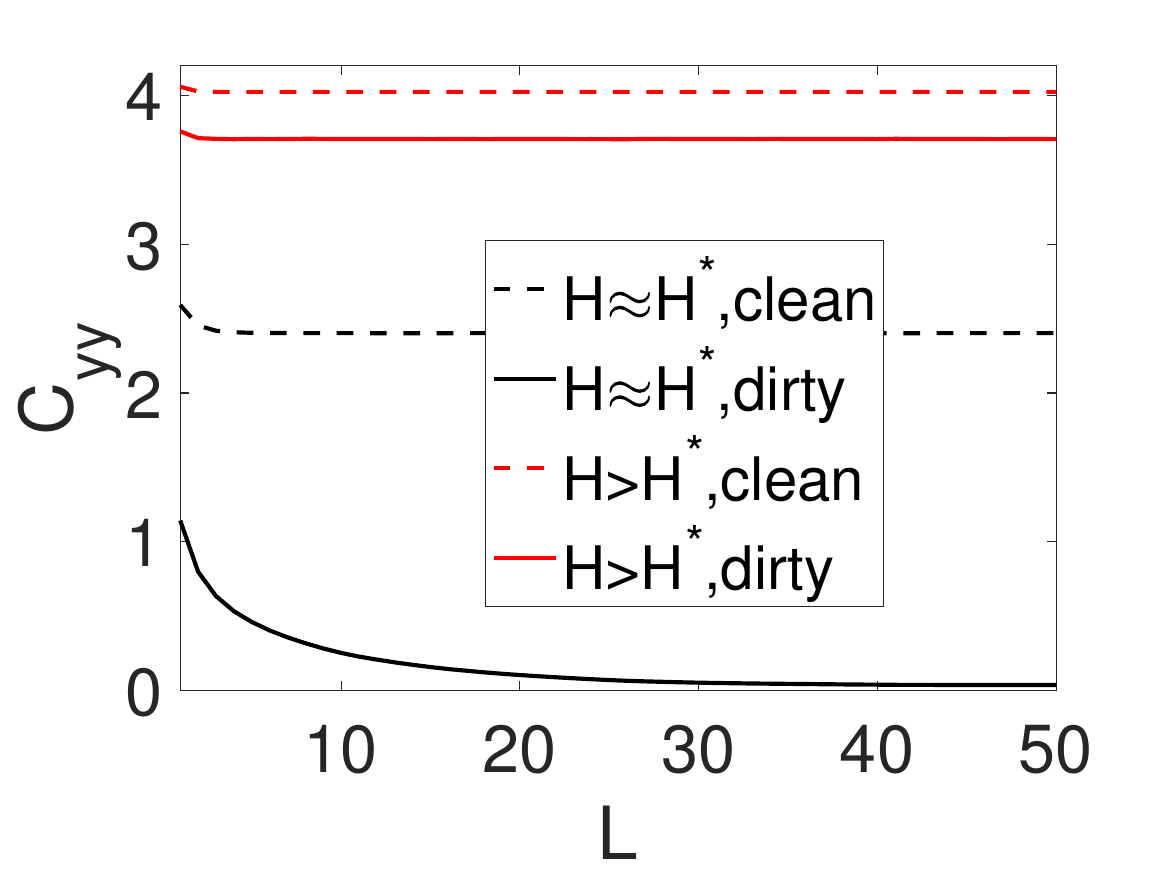}
		\caption{Correlation function of the dominant order. Black lines are at field $H\approx{H^*}$, where disorder destroys long-range ordering. Red lines are at the same temperature but higher field, where disorder weakly affects the correlation function. Reentrant superconductivity, therefore, appears as magnetic increases. }
		\label{f3}
	\end{figure}
	
	$L_x\times{L_y}\times{L_z}=100\times100\times2$ square lattice with periodic boundary condition is considered. In order to describe the long-range ordering, we compute the in-plane correlation function along the diagonal direction:
	\begin{equation}
		\begin{split}
			C_{A,B}(L)=\sum_{\bf r}\langle\Delta_{A,B}({\bf r})\Delta^*_{A,B}({\bf r}+(L,L,0))\rangle, 
		\end{split}
	\end{equation}
	where $\langle...\rangle$ is the thermal average in Monte Carlo simulation.

	For the square point, we take $\overline{u_A}=3.97$ and $\overline{u_B}=4$. { The similar value of $u_A$ and $u_B$ ensures that this point is close to $H=H^*$.} For the triangle point, we take $\overline{u_A}=2$ and $\overline{u_B}=4$, where we assumed that the magnetic field strongly suppresses order A but not order B. Other parameters are: $K_1=K_2=1$, $K_3=K_4=0.1$, $K_5=0.1$, $b_1=1/4$, $b_2=0$, $b_3=1$ and $\Delta{u}=2$. $\beta\equiv1/T=0.37$ is used in the Monte-Carlo simulation. 5000000 sweeps with Metropolis algorithm is applied. The system size is already sufficiently large, such that different disorder realizations have a small difference in the correlation function.
	
	The results are summarized in Fig.\ref{f3}. For the square point ($H\approx{}H^*$, black lines), disorder destroys long-range ordering. For the triangle point ($H>{}H^*$, red lines), the disorder only weakly reduces the correlation function. Superconductivity, therefore, reappears as the magnetic field increases.

	\subsection{purely orbital effect}
	We now consider the phase transition driven by a purely orbital effect.  In Landau gauge $\vec A = \left( Hy, 0 \right)$, the free energy density is
	\begin{equation}
		\begin{split}
			f({\bf r})&=-\frac{u_A({\bf r})}{2}|\Delta_A|^2-\frac{u_B({\bf r})}{2}|\Delta_B|^2\\
			&+b_1(|\Delta_A|^2+|\Delta_B|^2)^2+b_3|\Delta_A|^2|\Delta_B|^2\\
			&+\frac{K_A}{2}(|(\partial_x-iA_x)\Delta_A|^2+|\partial_y\Delta_A|^2)\\
			&+\frac{K_B}{2}(|(\partial_x-iA_x)\Delta_B|^2+|\partial_y\Delta_B|^2)
		\end{split}
	\end{equation}
	
	We consider a toy model on a finite 2-dimensional plane. The presence of the third dimensionality hardly affects our conclusions and we neglect it below for simplicity.   The assumption that phase A occurs at small fields, whereas phase B occurs at higher fields (i.e. has a higher upper critical field or shorter correlation length), requires $\overline{u_A}>\overline{u_B}$, and $K_A>K_B$.   
	%We neglect the anisotropy of the kinetic terms within the x-y plane for simplicity.  
	\begin{equation}
		\begin{split}
			&u_A({\bf r})=\overline{u_A}+\delta{u_A}({\bf r})\\
			&u_B({\bf r})=\overline{u_B}+\delta{u_B}({\bf r}).
		\end{split}
	\end{equation} 
	%	, with $\overline{u_A}>\overline{u_B}$ such that order A is dominant under zero field. $K_A>K_B$ is required for the orbital-driven transition.
	
	We study a system on a $L\times{L}=100\times100$ square lattice with periodic boundary conditions. For the square point in Fig.\ref{f1}, we take $H=7\times\frac{2\pi}{L}$. For the triangle point, we take $H=8\times\frac{2\pi}{L}$. { Note that the periodic boundary condition restricts the choice of the magnetic field strength. In numerical calculations, specific gauge may further restrict the choice. For the Landau gauge, $H$ needs to be a multiple of $\frac{2\pi}{L}$.} Other parameters are: $\overline{u_A}=1.444$, $\overline{u_B}=1$, $K_A=3$, $K_B=1$, $b_1=1/4$, $b_3=1$ and $\Delta{u}=0.4$. $\beta\equiv1/T=40$ is used in the Monte-Carlo simulation. { Proper value of $u_A$ is tuned such that the square point is close to $H=H^*$. This tuning is unnecessary, if one instead wants to scan over multiple magnetic fields in a much larger system. } $2\times{10^8}$ sweeps with Metropolis algorithm and over-relaxation algorithm are applied. The result is averaged over 20 disordered samples.

		\begin{figure}[h]
		\centering
		\includegraphics[width=6cm]{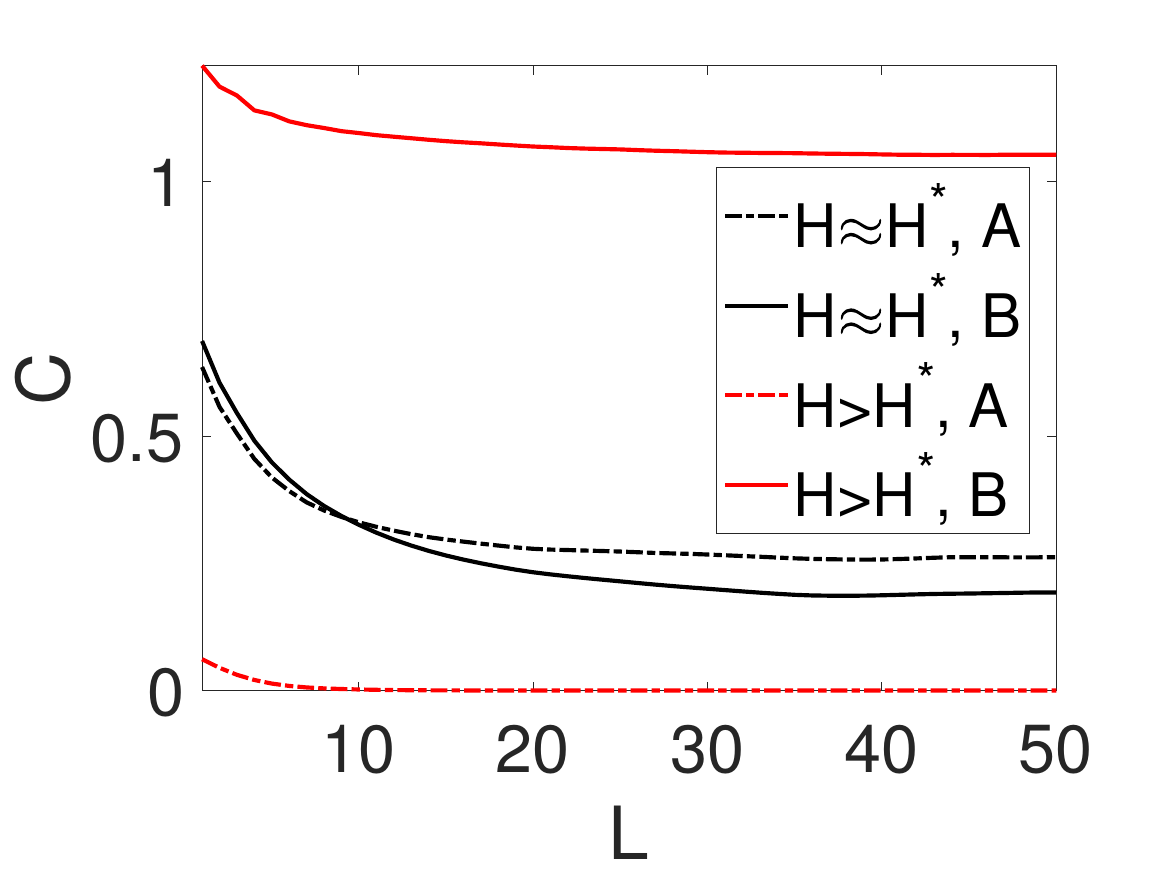}
		\caption{Correlation function of the dominant order (solid lines) and subdominant order (dash-dotted lines). Black lines are at field $H\approx{H^*}$, while red lines are at the same temperature but higher field. By increasing magnetic field, the correlation function at long distance of the dominant order increases by near an order of magnitude.}
		\label{f4}
	\end{figure}
	
	We compute the pair correlation function along the diagonal direction:
	\begin{equation}
		\begin{split}
			C_{A,B}(L)=\overline{\sum_{ x,y}|\langle\Delta_{A,B}(x,y)\Delta^*_{A,B}(x+L,y+L\rangle|}
		\end{split}\label{eq6}
	\end{equation}
	Note that the absolute magnitude is taken before adding up correlators, since correlators are complex when including the vector potential. The final overline denotes disorder-average. The results are summarized in Fig.\ref{f4}. For $H\approx{}H^*$ (black solid line), the correlation function is small due to imhomogeneity. By increasing magnetic field to $H>{}H^*$(red solid line), the correlation function of superconducting order parameters increases dramatically.  The correlation function of the subdominant order is also included as dashed-dotted lines. 
	
	{ Due to the absolute value in Eq.\ref{eq6}, the resulted correlation function cannot be used to distinguish the ordered and disordered phase. Instead, we now compute the superfluid stiffness along the Y-direction:}
	\begin{equation}
		\begin{split}
			&J^s_{A,B}=\overline{\frac{1}{L^2}\langle{E_Y}\rangle-\frac{\beta}{L^2}(\langle{I_Y^2}\rangle-\langle{I_Y}\rangle^2)}\\
			&E_Y=\frac{K_{A,B}}{2}\sum_{ x,y}(\Delta_{A,B}^*(x,y)\Delta_{A,B}(x,y+1)+c.c.)\\
			&I_Y=\frac{K_{A,B}}{2}\sum_{ x,y}(i\Delta_{A,B}^*(x,y)\Delta_{A,B}(x,y+1)+c.c.)
		\end{split}
	\end{equation}
	 For $H\approx{H^*}$, we get $J^s_A=0.03\pm0.05$ and $J^s_B=0.01\pm0.02$. For $H>{H^*}$, we get $J^s_A=0.0002\pm0.004$ and $J^s_B=0.11\pm0.03$, which shows the re-entrant superconductivity. The average and standard deviation is computed from 20 disordered samples.

	The spatial dependence of the order parameter can be found in the appendix.

	\section{discussion}\label{S4}
	In this work, we discussed the effect of inhomogeneity near a field-driven first-order transition. Inhomogeneity in UTe$_2$ has been observed in NMR \cite{nakamine2021} and specific heat experiments \cite{thomas2021}. Typically, the recent specific heat measurement observes different signals from different parts of a sample, which directly illustrates the importance of inhomogeneity. 
	
	In the above sections, we have assumed that the chiral phase is not relevant near $H=H^*$. The proposal of chiral phase \cite{hayes2020} at $H=H^*$ is supported by two experiments: (1)Kerr effect where time-reversal symmetry is found to be broken; (2) split transitions in specific heat, which are strictly required for the chiral phase in the orthorhombic system. However, the split transitions are observed in some samples \cite{hayes2020, thomas2021, rosa2021};  ], while other samples show a single transition \cite{ran2019,aoki2019,metz2019,thomas2021, rosa2021}, which puts the existence of the chiral phase into question.
	The proposed chiral phase has an intrinsic magnetization along the c-axis, which is perpendicular to the field $H_y$. Therefore, the chiral phase is always suppressed by the field. 
	In Landau theory, there are two mechanisms to suppress the chiral phase. One is through the suppression of the individual order parameters of A and B, as we analyzed in the previous sections. The second way is a direct suppression of the magnetization:
	\begin{equation}
		\begin{split}
			f=\alpha{}M_z^2H_y^2=\alpha(i\Delta_A^*\Delta_B+c.c.)^2H_y^2, 
		\end{split}
	\end{equation}
	with $\alpha>0$. In our analysis, we assumed that the suppression in the chiral phase is strong enough, such that it is irrelevant for the reentrant superconductivity. This assumption requires a sufficiently large $\alpha$. For smaller $\alpha$, the mean-field phase diagram is shown in the bottom panel. Our analysis on reentrant superconductivity still applies to the first-order transition line between the A and B phases. 
	
	\begin{figure}[h]
		\centering
		\includegraphics[width=6.4cm]{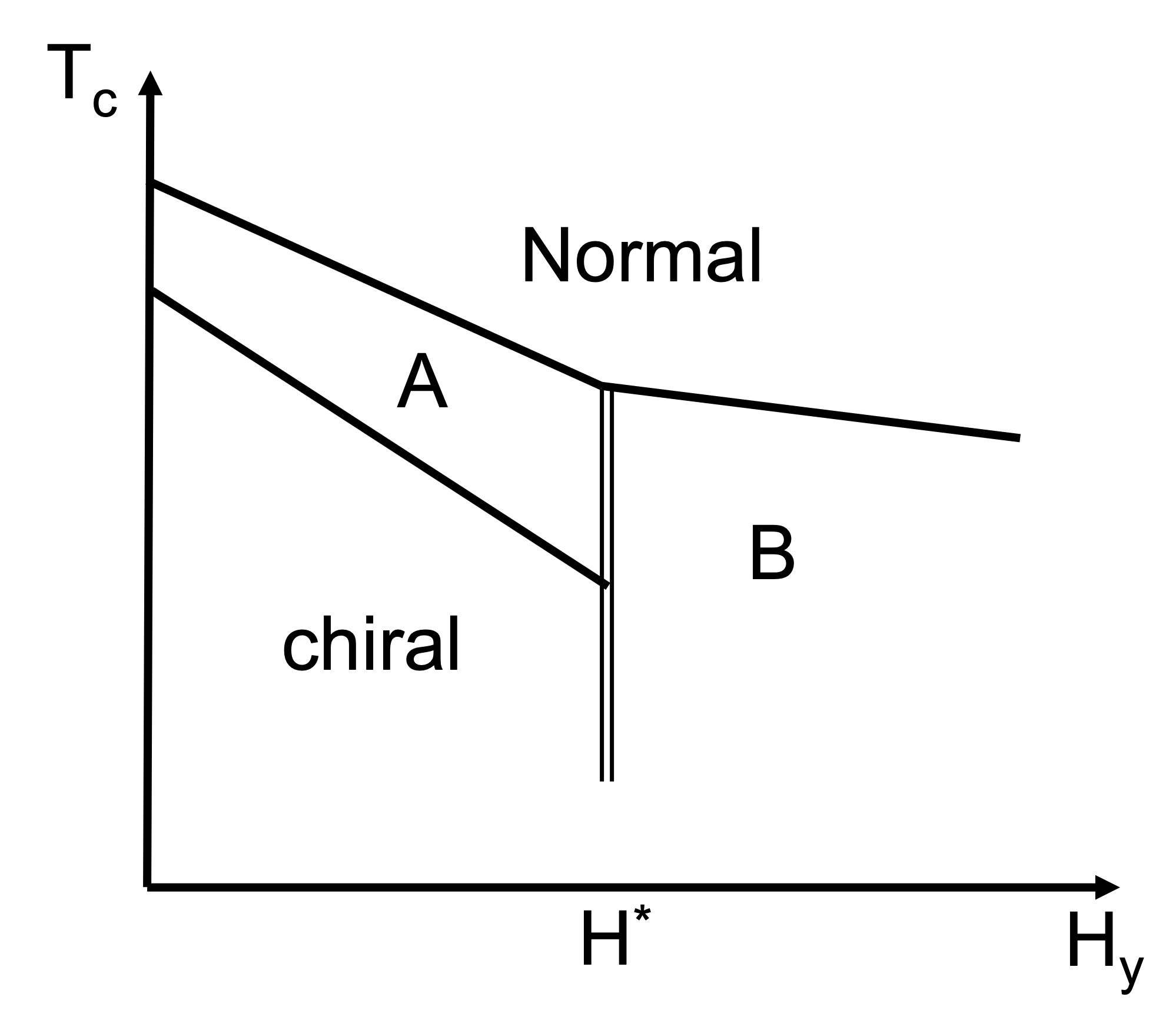}
		\caption{Mean-field phase diagram in the clean system, if the chiral phase is relevant. Single-line denotes continuous phase transitions, while double-line denotes first-order phase transitions.}
		\label{f5}
	\end{figure}
	
	With disorder, we proposed three possible fates for the field-driven transition. If the two phases belong to the irreducible representations for the chiral phase (Fig.\ref{f2}(a)), then there exist two continuous phase transitions at low temperature. Detecting the two continuous phase transitions (possibly in magnetocaloric effect) will further support the proposal of the chiral phase. The second scenario (change in vortex lattice structures Fig.\ref{f2}(b)) does not have phase transitions under disorder. But the local vortex lattice structures deep in the two phases can be distinguished in $\mu{SR}$ and small-angle neutron scattering. In the last scenario (same symmetry in the A\&B phase), there is only a first-order phase transition near the superconducting $T_c$. This phase transition could involve a sudden change in the position of the node, which could be observed in the penetration depth.  
	
	{ Our study has potential applications to other superconduting systems with a first-order transition. For example, CeRh$_2$As$_2$ \cite {Khim2021} is shown to have a first-order phase transition under magnetic field, between an even-parity superconducting phase and an odd-parity superconducting phase. Superconducting $T_c$ is slightly reduced near the phase transition.
		In the magic-angle twisted trilayer graphene \cite{cao2021}, field-driven reentrant superconductivity is also observed. Extra broadening of the superconducting transition is found near the minimal superconducting $T_c$, and the onset $T_c$ is a monotonically decreasing function of the magnetic field. These findings match with our study. }
	
	{\it Acknowledgments - }We thank S. Kivelson and J. Paglione for helpful discussions.  SR is supported  by the Department of Energy, Office of Basic Energy Sciences, Division of Materials Sciences and Engineering, under contract No. DE-AC02-76SF00515.

	\bibliography{citation}
	
	\clearpage
	\appendix
	\section{Spatial resolution}
	In this section, we would like to present the spatial dependence of the order parameters $|\langle{\Delta(x,y,z)}\rangle|$, for the orbital-effect driven transition.
	\begin{figure}[h]
		\centering
		\includegraphics[width=4cm]{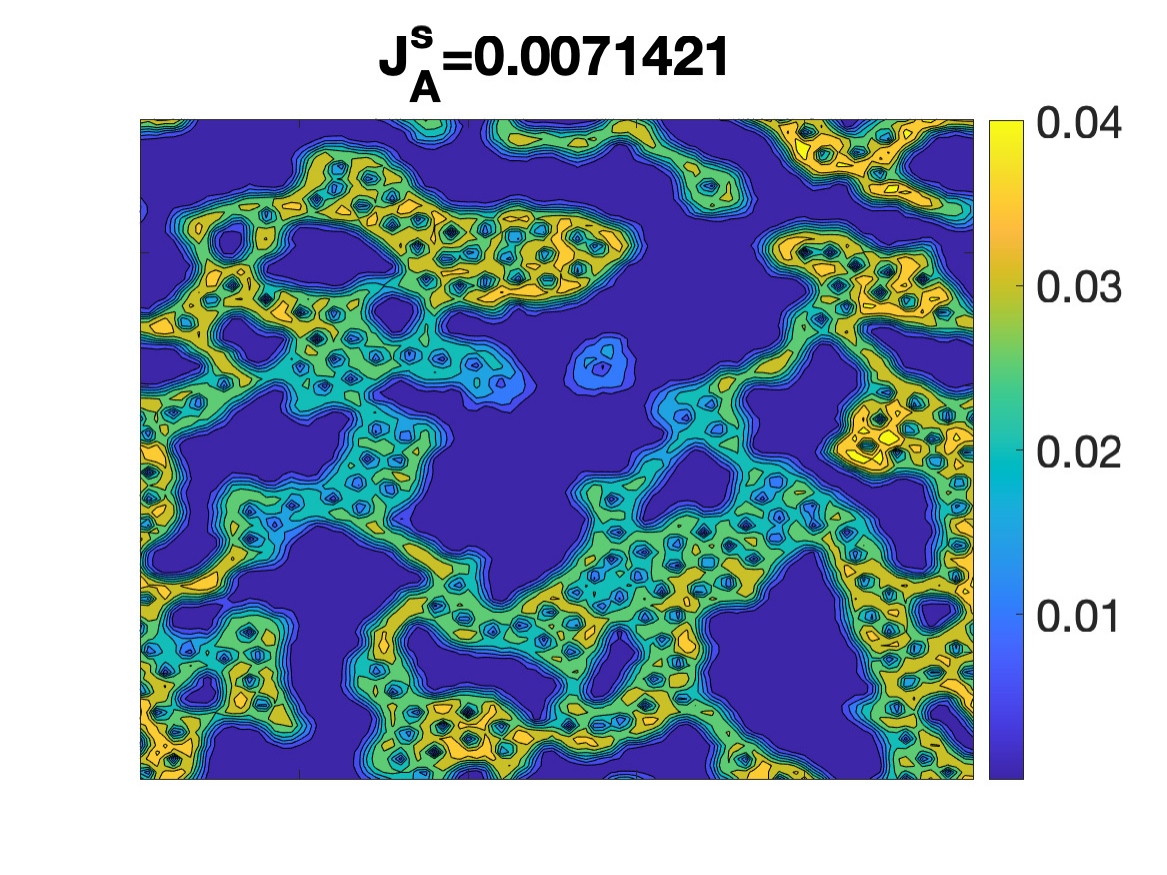}
		\includegraphics[width=4cm]{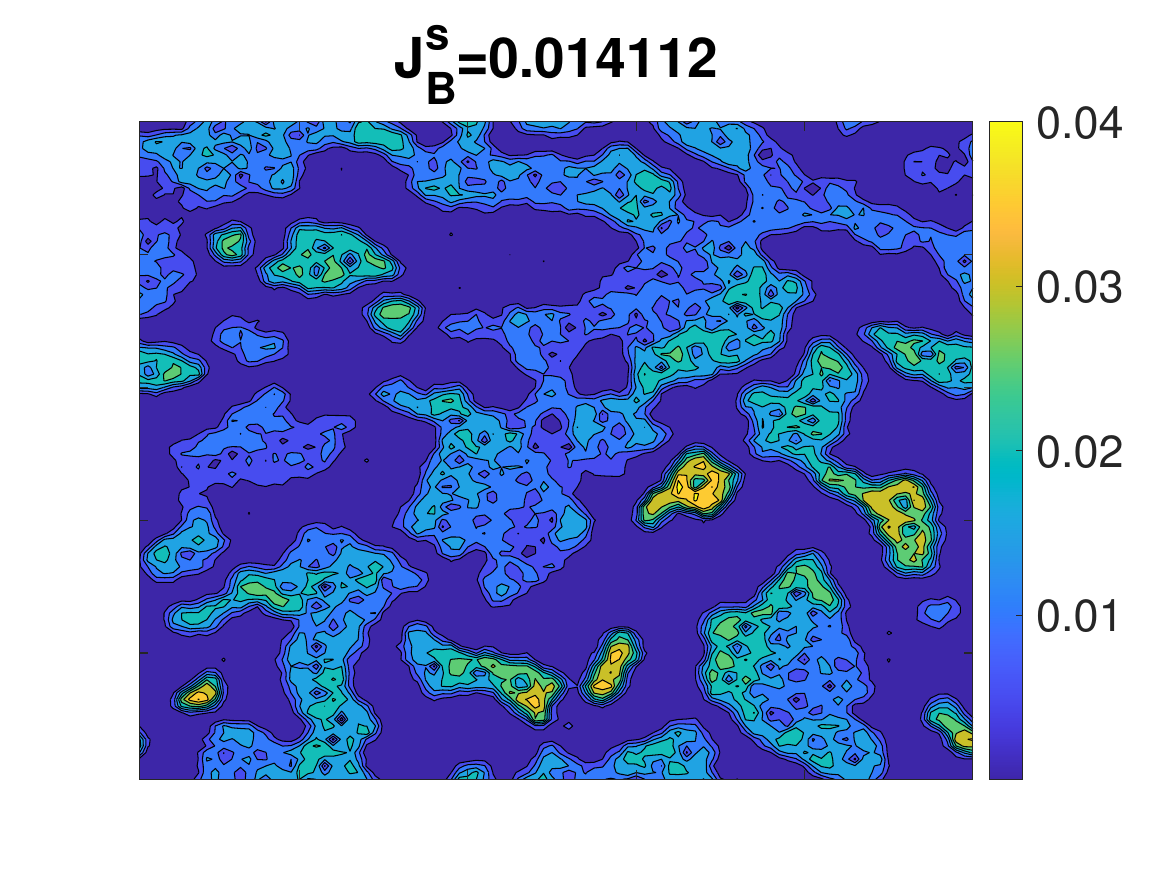}
		\includegraphics[width=4cm]{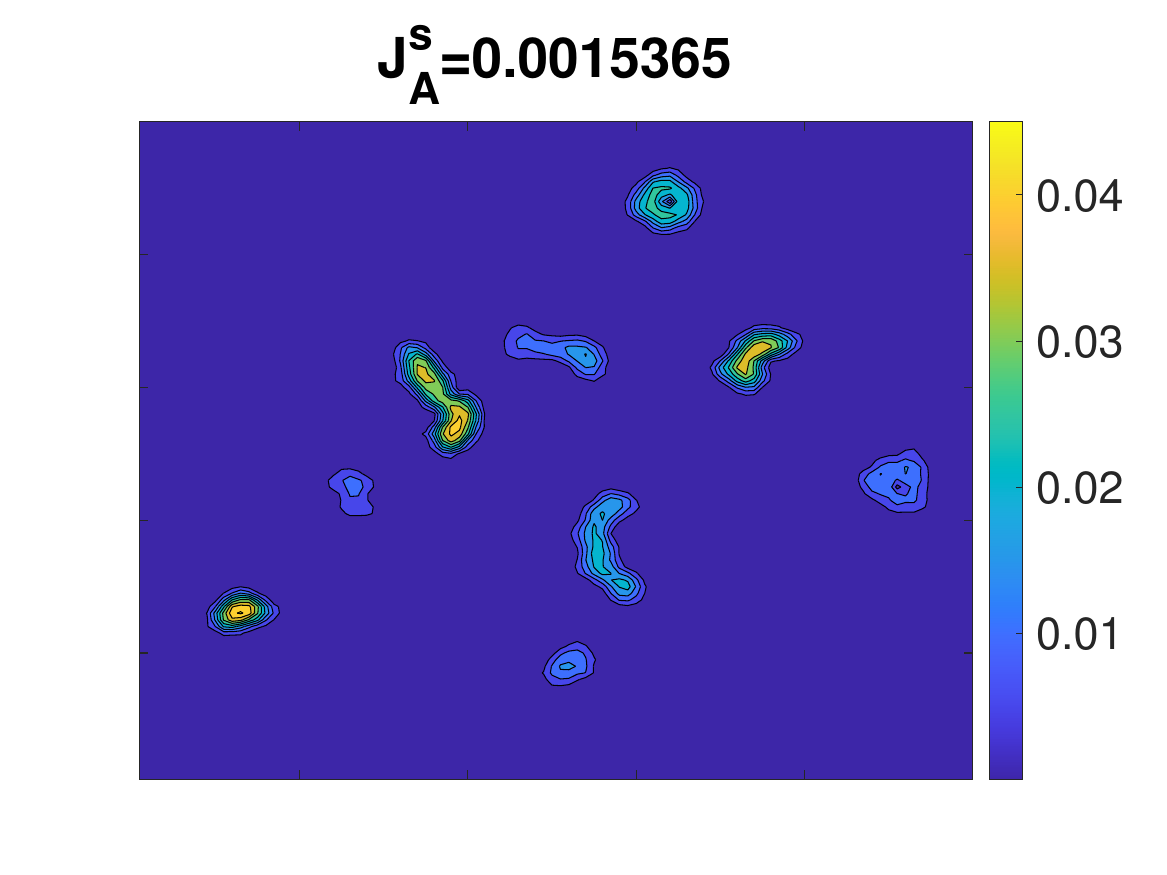}
		\includegraphics[width=4cm]{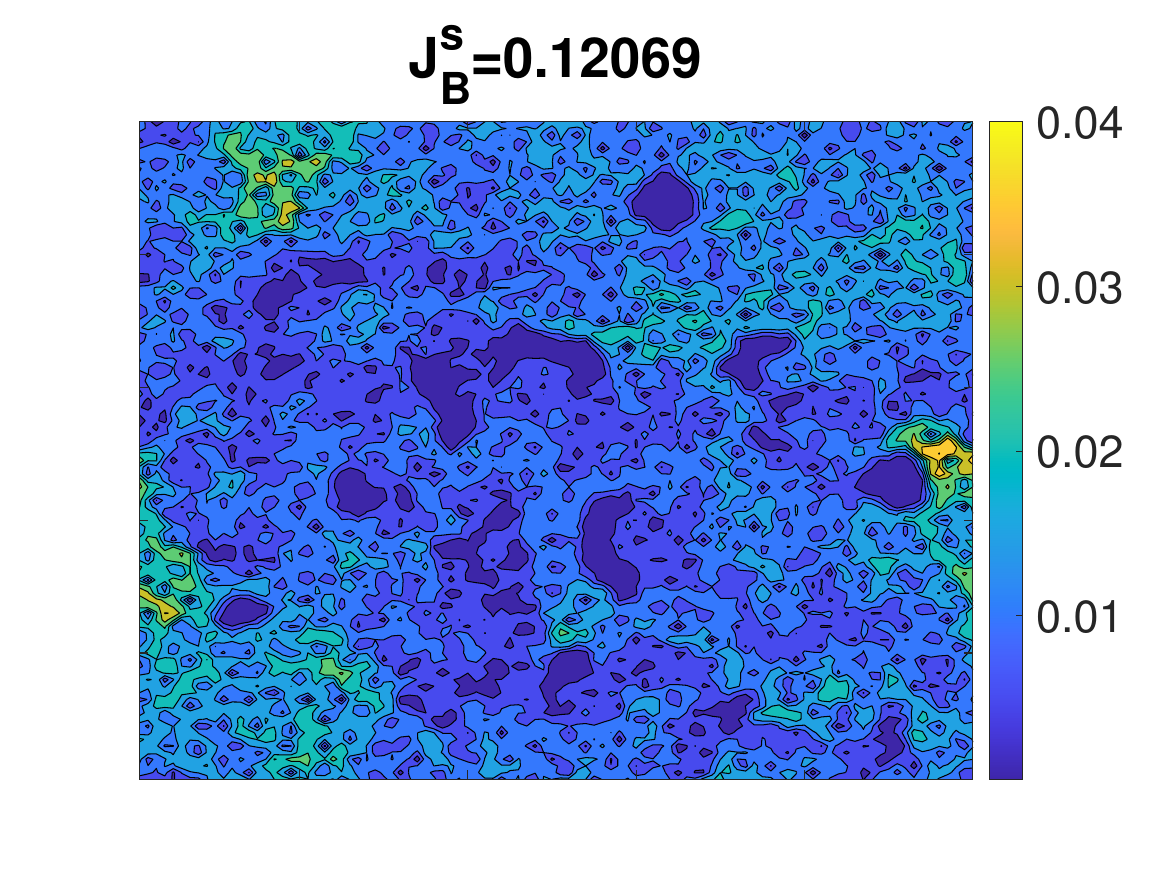}
		\caption{Magnitude of order parameters under orbital-effect-driven transition. The top two figures are at $H\approx{H^*}$, while the bottom two are at $H>{H^*}$. Increasing magnetic field strongly enhances the dominant order. The superfluid stiffness is included in the titles.}
	\end{figure}

\end{document}